\documentclass[prd,aps,twocolumn,preprintnumbers, showpacs, nofootinbib,superscriptaddress,notitlepage]{revtex4-1}
\usepackage{amsmath,graphicx,epsfig,amssymb}
\usepackage{inputenc}
\usepackage[usenames]{color}
\usepackage{ulem} 
\usepackage{bigstrut}
\usepackage{slashed}
\usepackage{multirow}
\usepackage{booktabs}
\usepackage{subcaption}
\usepackage{dsfont}
\usepackage{xcolor}
\usepackage{soul} 
\usepackage{tabularx}

\allowdisplaybreaks

\begin{document}



\title{Automated Extraction of Collins–Soper Kernel from Lattice QCD using An Autonomous AI Physicist System} 
\author{Jin-Xin Tan}
\affiliation{State Key Laboratory of Dark Matter Physics, Key Laboratory for Particle Astrophysics and Cosmology (MOE), Shanghai Key Laboratory for Particle Physics and Cosmology, School of Physics and Astronomy, Shanghai Jiao Tong University, Shanghai 200240, People's Republic of China} 
\author{Ting-Jia Miao}
\affiliation{School of Artificial Intelligence, Shanghai Jiao Tong University, Shanghai 200240, People's Republic of China} 
\affiliation{Zhiyuan College, Shanghai 200240, People's Republic of China} 
\author{Mu-Hua Zhang}
\affiliation{State Key Laboratory of Dark Matter Physics, Key Laboratory for Particle Astrophysics and Cosmology (MOE), Shanghai Key Laboratory for Particle Physics and Cosmology, School of Physics and Astronomy, Shanghai Jiao Tong University, Shanghai 200240, People's Republic of China} 
\author{Xiang-He Pang}
\affiliation{School of Artificial Intelligence, Shanghai Jiao Tong University, Shanghai 200240, People's Republic of China} 
\author{Ze-Xi Liu}
\affiliation{School of Artificial Intelligence, Shanghai Jiao Tong University, Shanghai 200240, People's Republic of China} 
\author{Lin-Feng Zhang}
\affiliation{DP Technology, Beijing 100190, People's Republic of China} 
\author{Si-Heng Chen}
\email{Corresponding author: sihengc@sjtu.edu.cn}
\affiliation{School of Artificial Intelligence, Shanghai Jiao Tong University, Shanghai 200240, People's Republic of China} 
\author{Wei Wang}
\email{Corresponding author: wei.wang@sjtu.edu.cn}
\affiliation{State Key Laboratory of Dark Matter Physics, Key Laboratory for Particle Astrophysics and Cosmology (MOE), Shanghai Key Laboratory for Particle Physics and Cosmology, School of Physics and Astronomy, Shanghai Jiao Tong University, Shanghai 200240, People's Republic of China}
\affiliation{Southern Center for Nuclear-Science Theory (SCNT), Institute of Modern Physics,
Chinese Academy of Sciences, Huizhou 516000, Guangdong Province, China}

\date{\today}

\begin{abstract}
We employ \textsc{PhysMaster}, an autonomous agentic AI system integrating theoretical reasoning, numerical computation, and exploitation strategies towards ultra-long horizon automation, to tackle long-standing challenges in non-perturbative lattice analyzes, including low signal-to-noise ratio at large transverse separation, complex systematic uncertainties, and labor-intensive manual workflows. Using the extraction of the CS kernel from quasi–transverse-momentum-dependent wave functions (quasi-TMDWFs) via large-momentum effective theory (LaMET) as a showcase, we demonstrate that \textsc{PhysMaster} automates high-dimensional fitting, renormalization, continuum–chiral extrapolation, and non-perturbative reconstruction in a fully autonomous manner. This framework drastically reduces the duration of the workflow from months to hours without compromising precision, stabilizes signals in the large-$b_\perp$ region to $1~\rm fm$, and produces results consistent with perturbative QCD and state-of-the-art traditional lattice calculations.   This work validates the effectiveness of physicist–AI collaboration for first-principles QCD research and establishes a generalizable, reproducible paradigm for automated studies of parton structure and other non-perturbative observables from lattice QCD. 
\end{abstract}
 
\maketitle

\section{Introduction}
\label{sec:intro}

Understanding the three-dimensional structure of nucleons is a central goal of modern hadronic physics.  Transverse-momentum-dependent (TMD) parton distribution functions (TMDPDFs)  encode both longitudinal motion and intrinsic transverse dynamics, enabling a comprehensive description of nucleon structure~\cite{Boussarie:2023izj}. The rapidity evolution of TMD observables—connecting measurements across different energy scales—is governed by the Collins-Soper (CS) kernel $K(b_{\perp}, \mu)$, a universal quantity that bridges soft and collinear dynamics in QCD \cite{Collins:1981uk,Collins:1981va,Collins:2011zzd}. Precise determination of the CS kernel's nonperturbative behavior is critical for accurate TMD factorization, resummation in QCD phenomenology, and reconciling data from experiments such as semi-inclusive deep-inelastic scattering (SIDIS) and Drell-Yan production.

Lattice QCD provides a first-principles framework for non-perturbative QCD calculations, and the large-momentum effective theory (LaMET) \cite{Ji:2013dva,Ji:2014gla,Ji:2020ect} allows the determination of the CS kernel through relating  Euclidean quasi-TMDWFs (computable on the lattice) to light-cone TMDs. Despite recent abundant advancements~\cite{Ebert:2018gzl, Shanahan:2020zxr, LatticeParton:2020uhz, Schlemmer:2021aij, Shanahan:2021tst, Li:2021wvl, LatticePartonLPC:2022eev, LatticePartonLPC:2023pdv, Shu:2023cot, Avkhadiev:2023poz, Liu:2024sqj, Avkhadiev:2024mgd, Alexandrou:2025xci,Tan:2025ofx},  there are still  several challenges. Firstly, nonlocal correlation functions decay exponentially with increasing transverse separation, and  can lead to poor signal-to-noise ratio (SNR) and unstable extractions. Secondly,  multiple extrapolations (infinite-momentum, continuum, chiral limits) and renormalization procedures introduce intricate systematic uncertainties that require careful handling. Moreover, extracting the CS kernel from quasi-TMDWFs involves fitting momentum-dependent correlators across multiple lattice ensembles, demanding efficient numerical techniques and model validation. Therefore,  lattice QCD calculations for exmaple in Ref.~\cite{Tan:2025ofx} remain labor-intensive, with the engineering aspects of data processing, fitting, and extrapolation often requiring great efforts. These limitations motivate the development of AI-driven tools to automate and optimize lattice QCD analyses.

In this work, we employ an autonomous agentic AI system \textsc{PhysMaster} \cite{Miao:2025sms} to address the challenges of CS kernel extraction. \textsc{PhysMaster} is an LLM-based autonomous agent designed for theoretical and computational physics research, integrating Monte Carlo Tree Search (MCTS) for long-horizon task navigation, a layered academic data universe (LANDAU) for reliable knowledge reuse, and seamless integration of theoretical reasoning with code execution.

By applying this framework to state-of-the-art lattice data \cite{Tan:2025ofx}, we demonstrate that \textsc{PhysMaster} reduces the labor-intensive engineering workflow from months to hours while maintaining precision; achieves consistency traditional lattice results; improves SNR in the large-$b_{\perp}$ region ($b_{\perp} > 0.5$ fm) compared to direct extraction methods through imposing physics-inspired constraints. 
This study not only advances the state-of-the-art in CS kernel extraction but also establishes a paradigm for AI-automated lattice QCD analyses   of nonperturbative observables in QCD.


\section{Theoretical and Methodological Framework}
\label{sec:theory}

\subsection{Determining  CS kernel within LaMET}
\label{subsec:lamet}

Within the LaMET framework, the quasi-TMDWF $\tilde{f}(x, b_{\perp}, \mu, \zeta_{z})$ for a meson boosted to large longitudinal momentum $P^{z}$ is defined as \cite{Ji:2020ect, Ji:2022ezo}:
\begin{eqnarray}
\tilde{f}\left(x, b_{\perp}, \mu, \zeta_{z}\right) &=& \lim_{L \to \infty} \int \frac{d z}{2 \pi} e^{i\left(x-\frac{1}{2}\right) z P^{z}}  \nonumber\\
&&\times \frac{\tilde{\Phi}^{0}\left(z, b_{\perp}, P^{z}, L\right)}{Z_{O}(1/a, \mu) \sqrt{Z_{E}\left(2L+z, b_{\perp}\right)}}, 
\end{eqnarray}
where $\tilde{\Phi}^{0}$ is the bare nonlocal matrix element constructed from gauge-invariant operators with staple-shaped Wilson links, $Z_{O}$ is the renormalization factor for logarithmic divergences, and $Z_{E}$ (derived from Wilson loops) cancels linear divergences from the Wilson links. The rapidity scale $\zeta_{z} = (2xP^{z})^{2}$ connects quasi-TMDWFs to physical light-cone TMDs.

The CS kernel governs the rapidity evolution of quasi-TMDWFs, with the relation between quasi-TMDWFs at different boost momenta $P_{1}^{z}$ and $P_{2}^{z}$ given by \cite{Ji:2020ect, LatticePartonLPC:2022eev}:
\begin{eqnarray}\label{NLOkernel}
&&K\left(b_{\perp}, \mu\right) = \frac{1}{\ln\left(P_{1}^{z}/P_{2}^{z}\right)}  \nonumber\\
&&\times  \ln\left[\frac{H\left(\zeta_{z2}, b_{\perp}, \mu\right) \tilde{f}\left(x, b_{\perp}, \mu, \zeta_{z1}\right)}{H\left(\zeta_{z1}, b_{\perp}, \mu\right) \tilde{f}\left(x, b_{\perp}, \mu, \zeta_{z2}\right)}\right] + \mathcal{O}\left(\frac{\Lambda_{\text{QCD}}^{2}}{\zeta_{z}}\right). \notag\\
\end{eqnarray}
Here, $H$ is the perturbative matching kernel, and we adopt the $b_{\perp}$-unexpanded next-to-leading order (uNLO) matching to avoid small-$b_{\perp}$ expansion artifacts \cite{Avkhadiev:2023poz}. The CS kernel is extracted by extrapolating to the infinite-momentum limit ($P^{z} \to \infty$), suppressing power corrections of $\mathcal{O}(1/(P^{z})^{2})$~\cite{Ji:2019ewn, Ji:2021znw, LatticePartonCollaborationLPC:2022myp, LatticeParton:2024mxp, Avkhadiev:2024mgd,LatticeParton:2025eui}. This approach has been widely applied to extract the CS kernel from lattice QCD~\cite{Ebert:2018gzl, Shanahan:2020zxr, LatticeParton:2020uhz, Schlemmer:2021aij, Shanahan:2021tst, Li:2021wvl, LatticePartonLPC:2022eev, LatticePartonLPC:2023pdv, Shu:2023cot, Avkhadiev:2023poz, Liu:2024sqj, Avkhadiev:2024mgd, Alexandrou:2025xci,Tan:2025ofx}. 

\subsection{Agent-automated Reconstruction Strategy}
\label{subsec:agent}


\begin{figure}[h!]
\centering
\includegraphics[width=0.9\linewidth]{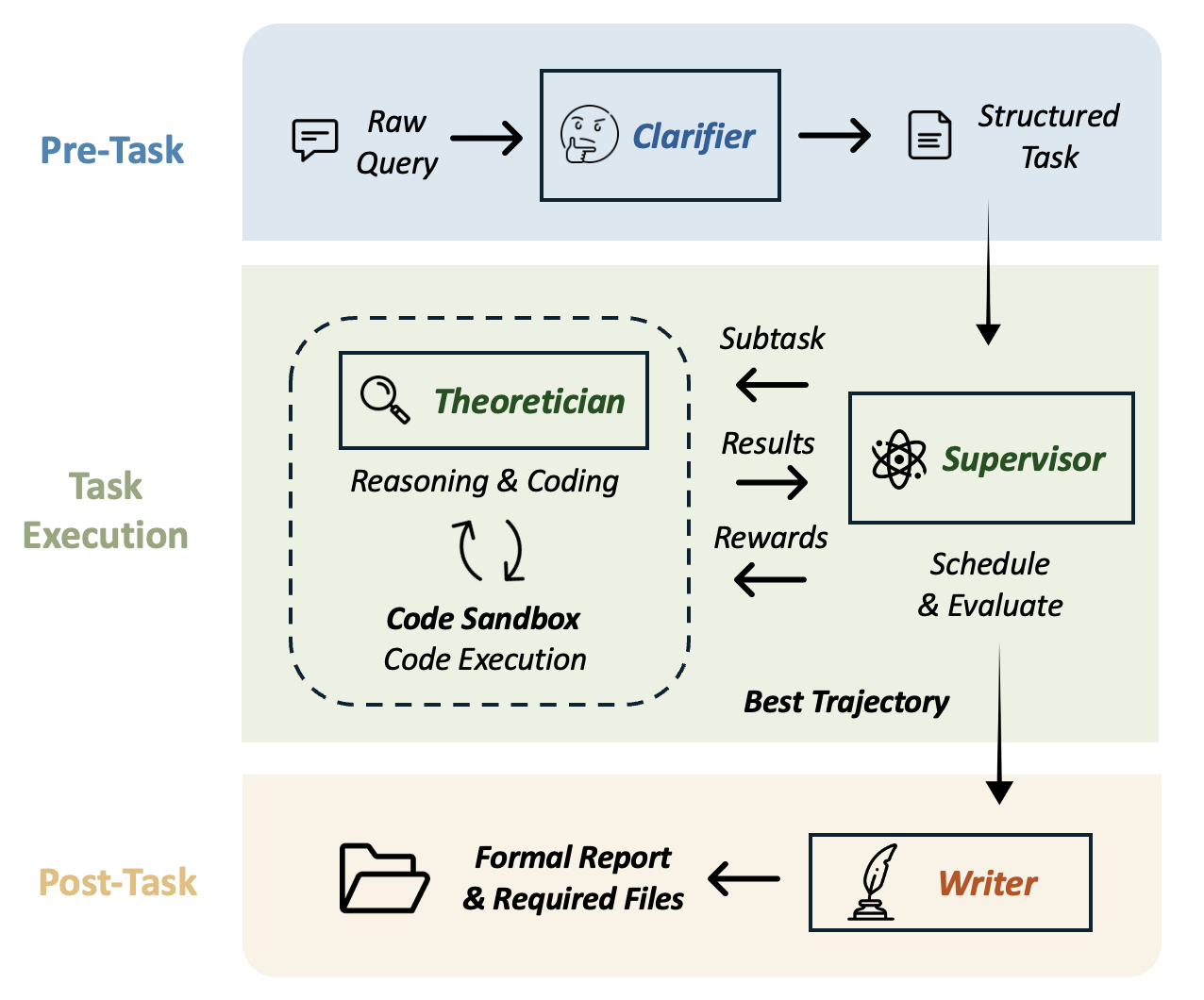}
\caption{\textsc{PhysMaster} workflow for CS kernel extraction. The framework integrates literature retrieval, MCTS-driven exploration, hierarchical agent collaboration, and result validation.}
\label{fig:workflow}
\end{figure}

\textsc{PhysMaster} allows  an end-to-end  extraction of physical observables from lattice QCD via a three-stage workflow as shown in Fig.~\ref{fig:workflow}, addressing the limitations of traditional manual analyses:
\begin{itemize}
\item Pre-task stage: This stage is further divided into  query clarification and literature retrieval and local library construction. More explicitly,  \textsc{PhysMaster} first decomposes the  extraction task for instance the determination of CS kernel into subtasks (correlator fitting, renormalization, extrapolation, kernel extraction) and identifies key constraints, for instance symmetry requirements and perturbative behavior at small $b_{\perp}$). Then \textsc{PhysMaster} builds a task-specific knowledge base integrating prior results from traditional lattice QCD \cite{Tan:2025ofx}, LaMET formalism \cite{Ji:2013dva, Ji:2014gla, Ji:2020ect, Ji:2022ezo}, and perturbative QCD predictions \cite{Li:2016ctv, Moch:2017uml, Vladimirov:2016dll, Moult:2022xzt, Duhr:2022yyp}.

\item Task execution stage:  This stage includes MCTS-driven exploration, hierarchical agent collaboration and  parametrization and reconstruction.   \textsc{PhysMaster} uses Monte Carlo Tree Search to navigate high-dimensional fitting and model selection, balancing exploration of parametrization forms with exploitation of physics constraints. Secondly, a ``supervisor" agent manages progress and evaluates results, while a ``theoretician" agent performs analytical derivations, code execution, and numerical fitting. At the end of this stage,   \textsc{PhysMaster} implements a physics-motivated parametrization of quasi-TMDWFs to stabilize large-$b_{\perp}$ signals.

\item Post-task stage: At this stage,  \textsc{PhysMaster} validates the results  by  comparing  extracted CS kernels with traditional lattice results and perturbative predictions, and  quantifying  systematic uncertainties, and generates structured reports.
\end{itemize}
More details of \textsc{PhysMaster} can be found in the appendix and Ref. \cite{Miao:2025sms}.

\section{Lattice QCD Calculation and \textsc{PhysMaster} Results}
\label{sec:results}
 
\subsection{Long-chain Analysis as Automated Lattice Workflow}

\begin{figure*}[ht]
\centering
\includegraphics[width=0.32\linewidth]{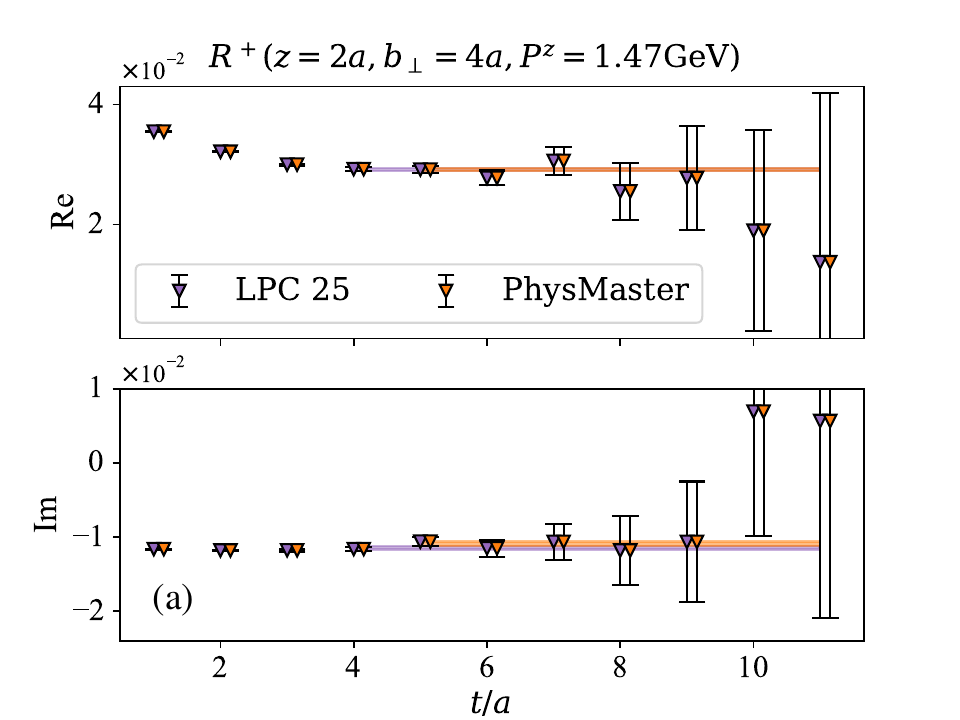}
\includegraphics[width=0.32\linewidth]{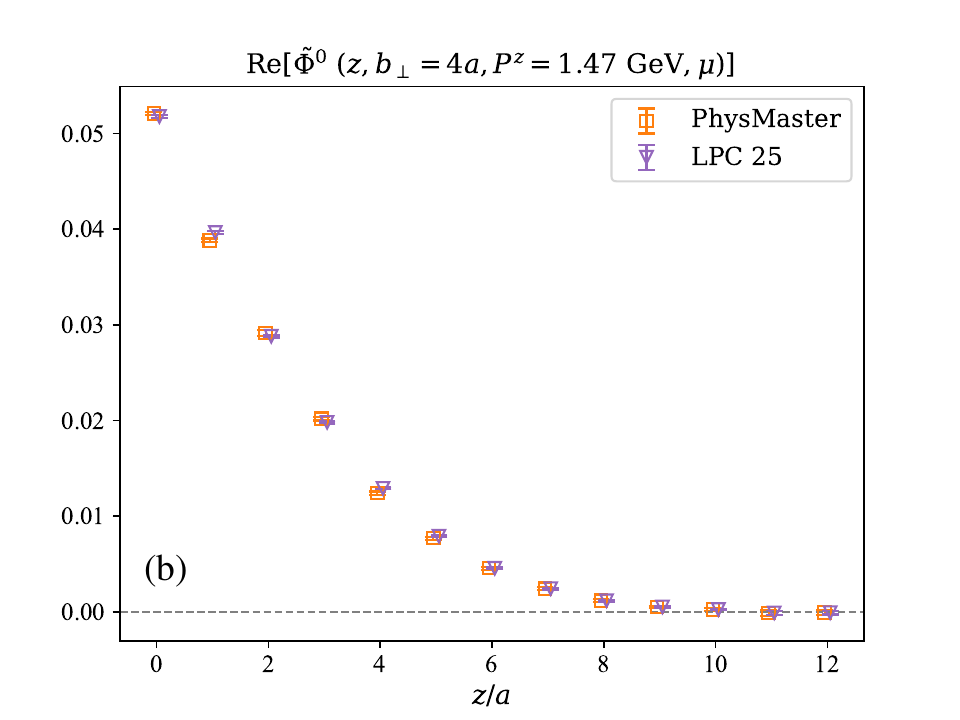}
\includegraphics[width=0.32\linewidth]{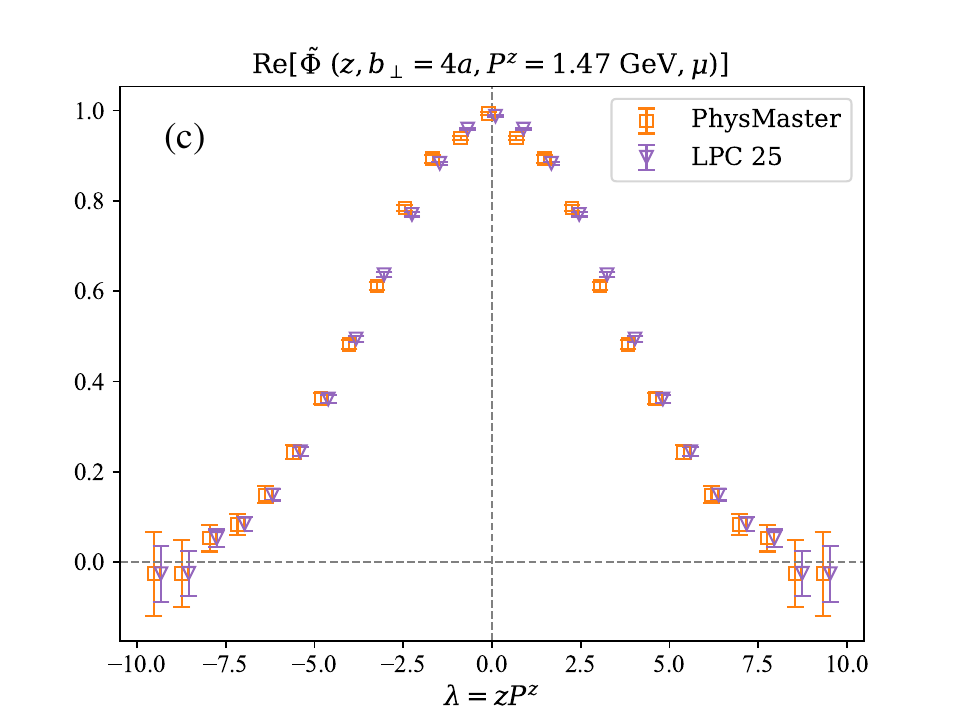}    
\includegraphics[width=0.32\linewidth]{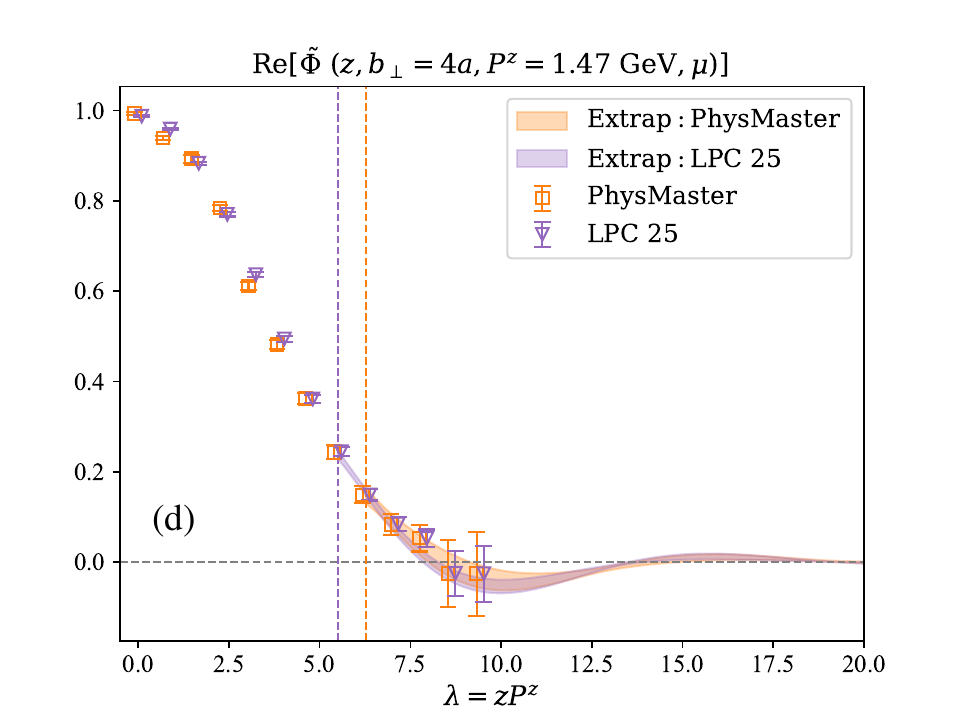}
\includegraphics[width=0.32\linewidth]{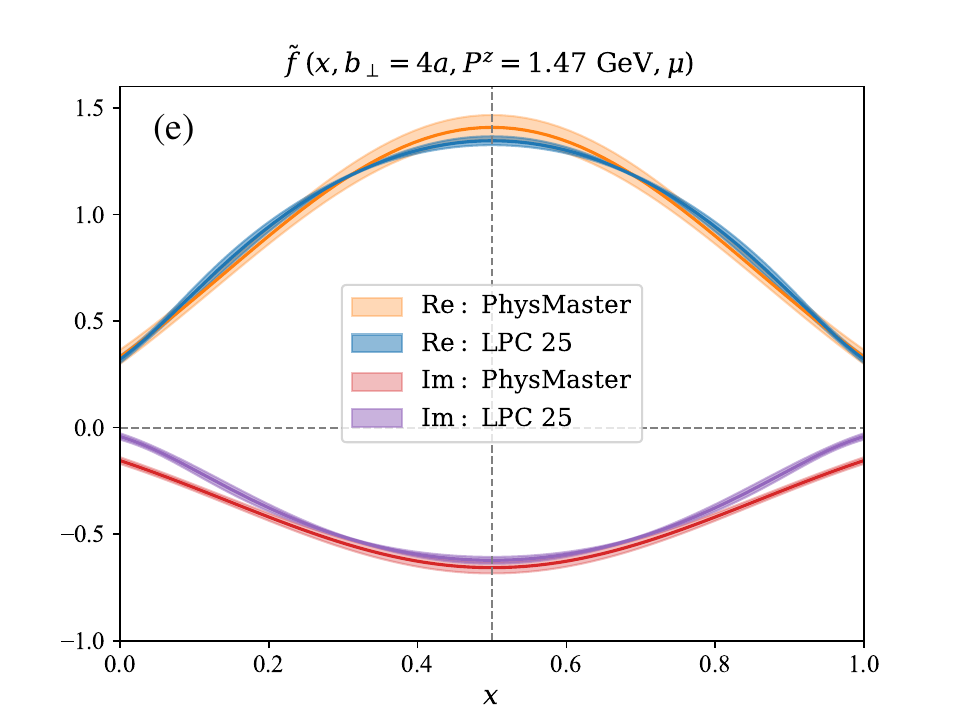}    
\includegraphics[width=0.32\linewidth]{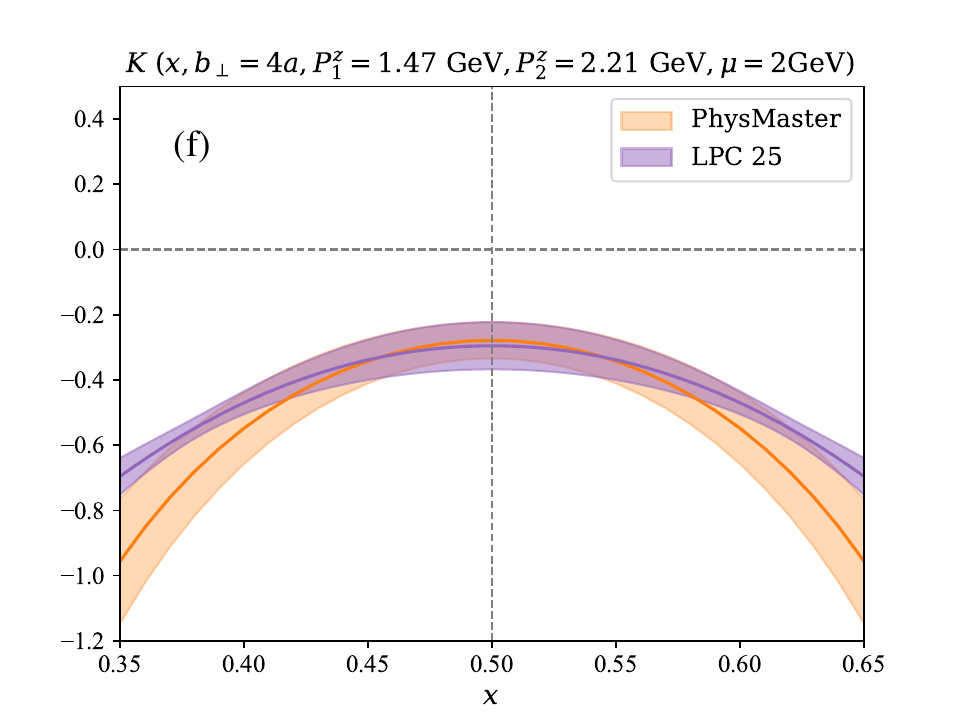}
\caption{ Automated long‑chain analysis workflow executed by \textsc{PhysMaster} for CS kernel extraction. Starting from raw lattice two‑point correlators and Wilson loop data, the agent performs correlator ratioing, plateau fitting, renormalization, tail stabilization, Fourier transformation, and final CS kernel extraction in a fully automated manner. }
\label{fig:long-chain}
\end{figure*}

In this section we summarize the full lattice-QCD workflow executed by the \textsc{PhysMaster} to extract the CS kernel from quasi-TMD wave functions of the pion.  The system performs an end-to-end chain of operations, starting from raw Euclidean two-point correlators $C_2$ and Wilson-loop data provided by Ref.~\cite{Tan:2025ofx}, and ending with a statistically and systematically controlled lattice determination of $K(b_\perp,\mu)$ at fixed transverse separation.  An overview of the workflow is given in Fig.~\ref{fig:long-chain}, where  $b_{\perp} = 4a$ and $\mu = 2$ GeV are used. For illustrative purpose.  we choose the  C32P23 lattice ensembles.   


We begin by computing the ratio of nonlocal to local correlators to reduce the hadron state dependence and to remove the time dependence:
\begin{eqnarray}
R(z, b_{\perp}; t) = \frac{C_{2}(z, b_{\perp}; t)}{C_{2}(0,0;t)}. 
\end{eqnarray}
As shown in Fig.~(\ref{fig:long-chain}a), the agent automatically identifies the plateau region using log-effective-plateau diagnostics and performs either one-state or two-state fits to extract the bare quasi-TMDWF matrix element $\Phi_{\rm bare}(z, b_{\perp}) = \lim_{t \to \infty} R(z, b_{\perp}; t)$.


To remove linear and logarithmic divergences that bare matrix element suffers, \textsc{PhysMaster} applies the renormalization procedure:
\begin{eqnarray}
\tilde{\Phi}^{R}(z, b_{\perp}) = \frac{\Phi_{\rm bare}(z, b_{\perp})}{Z_O \sqrt{Z_E(z + 2L, b_{\perp})}}
\end{eqnarray}
where $Z_E$ is the Wilson-loop renormalization factor and $Z_O$ accounts for operator renormalization. The agent correctly identifies the $z + 2L$ entry for each matrix element and performs pointwise renormalization, resulting in significantly improved large-$z$ behavior compared to bare results as shown in Fig.~(\ref{fig:long-chain}b)  and (\ref{fig:long-chain}b).

At large $\lambda = zP^z$, lattice signals suffer exponential noise degradation. \textsc{PhysMaster} stabilizes the correlator tail using a physics-motivated parametrization~\cite{Ji:2020brr}:
\begin{eqnarray}
\tilde{\Phi}(\lambda, b_{\perp}) = \left[\frac{c_1}{(-i\lambda)^{n_1}} + e^{i\lambda} \frac{c_2}{(i\lambda)^{n_2}}\right] e^{-\lambda/\lambda_0}, 
\end{eqnarray}
where the fit parameters are constrained by Bayesian priors. The agent performs simultaneous fits to the real and imaginary parts over sliding windows of large $\lambda$, replacing noisy data with smoothly continued values shown in Fig.~(\ref{fig:long-chain}d).

To obtain the momentum-space distributions, the \textsc{PhysMaster} performs a Fourier transformation
\begin{eqnarray}
\tilde{f}(x, b_{\perp}, P^z) = \int \frac{d\lambda}{2\pi} e^{i(x - 1/2)\lambda} \tilde{\Phi}(\lambda, b_{\perp})
\end{eqnarray}
The results in Fig.~(\ref{fig:long-chain}e) show consistency with manual analyses, with minor differences in the real part attributed to different asymptotic continuation strategies—typically accounted for as systematic uncertainty.

Finally, \textsc{PhysMaster} extracts the CS kernel using the LaMET Eq.(\ref{NLOkernel}). The residual $x$ and $P^z$ dependence is removed by fitting the leading $\mathcal{O}(1/(P^{z})^{2})$ correction in the central region ($x \in [0.3, 0.7]$) and extrapolating to $P^{z} \to \infty$.

One can observe that despite of  the long workflow,  \textsc{PhysMaster} can predict the correct result for the CS kernel at $b_{\perp} = 4a$ and $\mu = 2$ GeV, where the central value aligns with benchmark data within uncertainties, confirming its reliability. It is necessary to notice that the full workflow of data processing, high‑dimensional fitting, continuum and chiral extrapolation is shortened from  months of manual work to a few hours. 


\subsection{$b_{\perp}$ Parametrization and Extracted CS Kernel}
\label{subsec:fit_results}

\begin{figure*}[ht]
\centering
\includegraphics[width=0.32\linewidth]{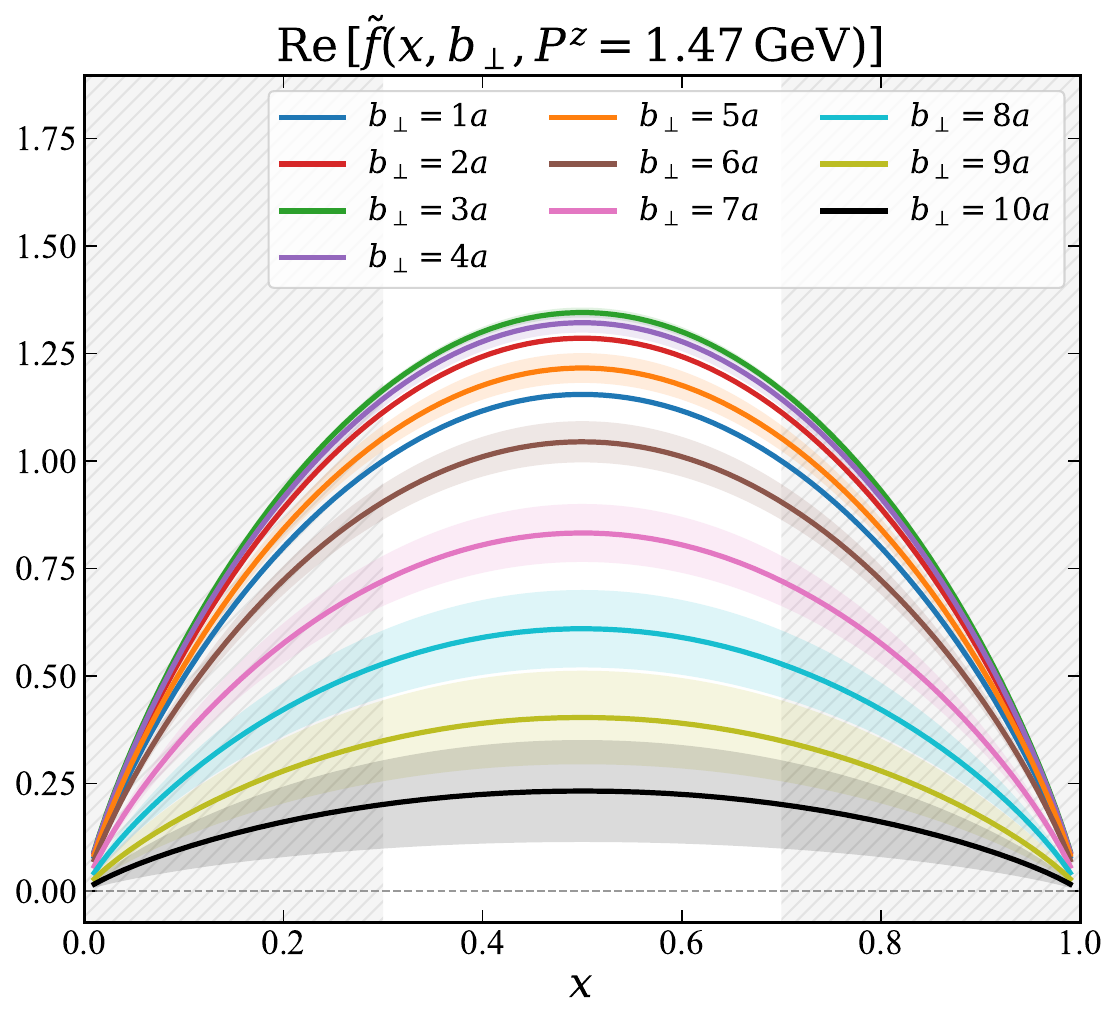}\hspace{1.cm}
\includegraphics[width=0.32\linewidth]{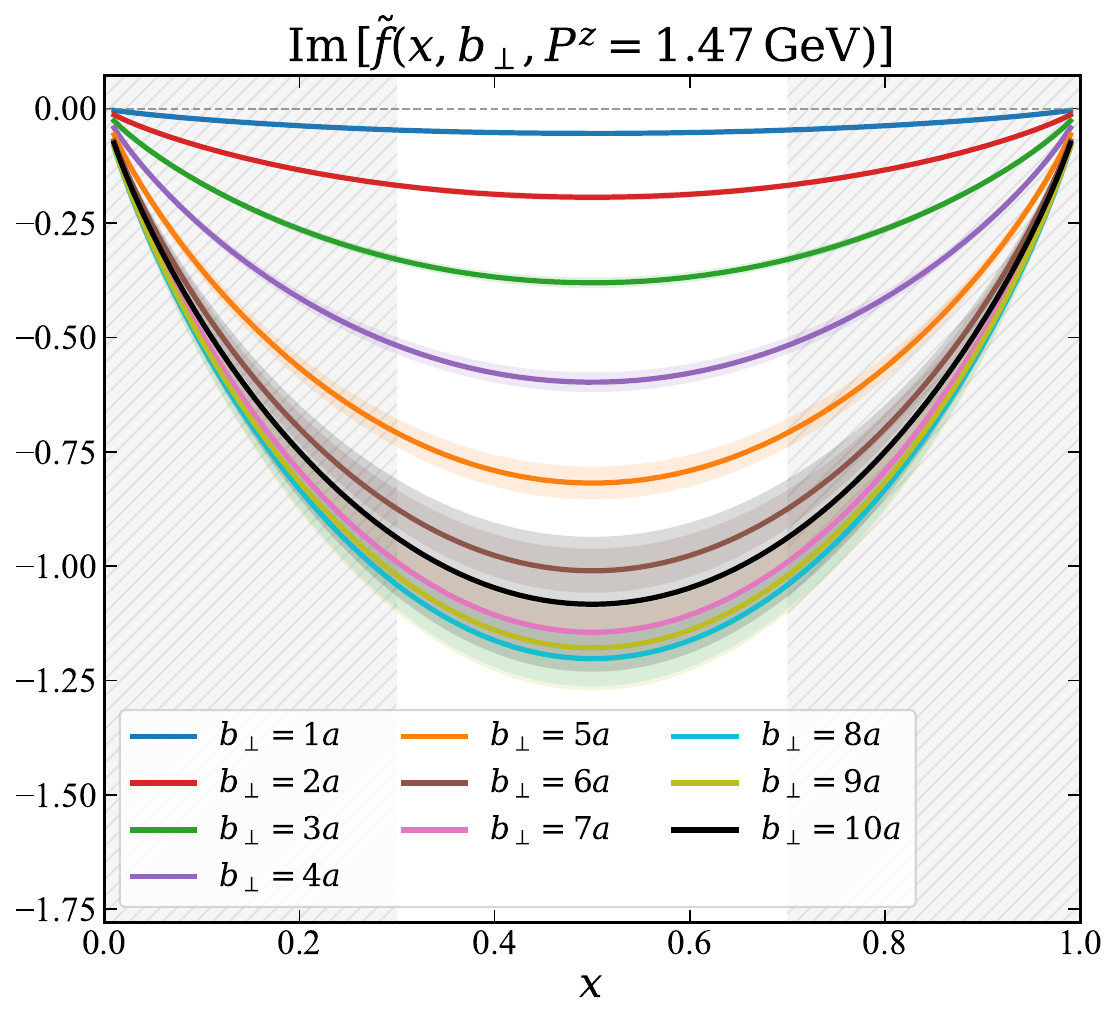}
\caption{Renormalized quasi‑TMD wave function (quasi‑TMDWF) at longitudinal momentum $P^z=1.47$ GeV. Real and imaginary parts are shown as functions of the momentum fraction $x$, demonstrating stable behavior enforced by \textsc{PhysMaster}’s physics‑motivated parametrization.  }
\label{fig:cs_kernel1}
\end{figure*}

\begin{figure}[htb]
\centering
\includegraphics[width=0.8\linewidth]{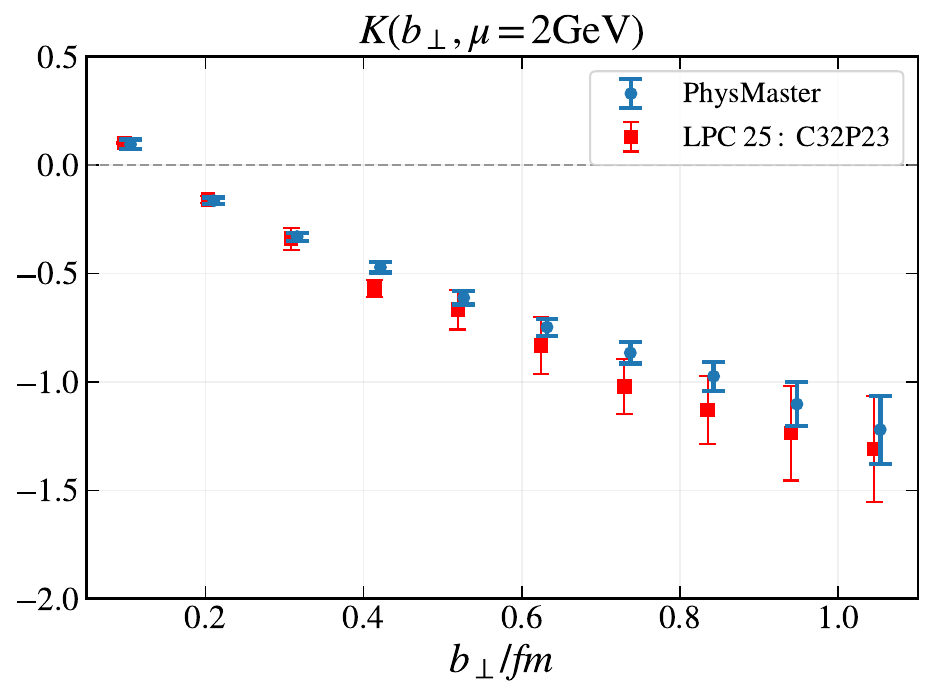}
\caption{Collins–Soper kernel $K(b_\perp, \mu=2{\rm GeV})$ extracted by \textsc{PhysMaster}.  Results provide robust  nonperturbative constraints up to $b_\perp \sim 1$ fm  with improved stability relative to traditional lattice extraction~\cite{Tan:2025ofx}.   } 
\label{fig:cs_kernel2}
\end{figure}

To address the signal loss at large $b_{\perp}$, \textsc{PhysMaster} can  parametrize the momentum-space quasi-TMDWF $\tilde{f}(x, b_{\perp}, P^{z})$ by imposing a form constrained by both perturbative QCD and lattice data. For illustrative purpose, we let it use: 
\begin{eqnarray}
\mathrm{Re}\,\tilde f(x,b_\perp,P^z) &= N\,x^{\alpha}(1-x)^{\alpha}\exp\!\left(-g_1\,b_\perp-g_2\,b_\perp^2\right) \notag\\
&\times \left(a_0+a_1\,b_\perp+a_2\,b_\perp^2\right), \\
\mathrm{Im}\,\tilde f(x,b_\perp,P^z) &= N\,x^{\alpha}(1-x)^{\alpha} \exp\!\left(-g_1\,b_\perp-g_2\,b_\perp^2\right) \notag\\
&\times \left(c_0+c_1\,b_\perp+c_2\,b_\perp^2\right). 
\end{eqnarray}
The polynomial term $x(1-x)$ captures the expected endpoint behavior of parton distributions (vanishing at $x=0$ and $x=1$). The exponential term  suppresses unphysical oscillations at large $b_{\perp}$, while the polynomial parametrizes other effects.

\textsc{PhysMaster} determines the parameters  by fitting to lattice quasi-TMDWF data in the intermediate $x$ range ($x \in [0.3, 0.7]$), where power corrections are minimized. The small-$b_{\perp}$ region ($b_{\perp} < 0.3$ fm) is further constrained by perturbative QCD predictions to reduce model dependence. The constrained  results are shown in Fig.~\ref{fig:cs_kernel1}.


The final Collins–Soper kernel $K(b_\perp, \mu=2{\rm GeV})$  extracted by \textsc{PhysMaster} is displayed in Fig.~\ref{fig:cs_kernel2}. The result is consistent with state-of-the-art lattice QCD determinations within uncertainties. At large $b_\perp$ up to 1fm, the physics-constrained parametrization improves stability compared to direct extractions, allowing robust nonperturbative constraints that are difficult to achieve in unconstrained fits. The entire fitting and extraction procedure is completed automatically, demonstrating that \textsc{PhysMaster} can reliably implement physically constrained lattice QCD analysis and produce publication-quality results without strong human supervision.


\section{Discussion and Prospect}
\label{sec:discussion}


The lattice determination of the CS kernel demonstrates that the automated \textsc{PhysMaster} framework delivers substantial improvements over conventional lattice QCD calculations in efficiency, objectivity, stability, and extensibility. The full workflow, including data processing, high-dimensional fitting, and continuum-chiral extrapolation, is reduced from months of manual effort to a few hours. Meanwhile, MCTS-driven model selection and objective fit validation minimize subjective choices in fitting ranges and parametrization forms, and physics-constrained reconstruction of quasi-TMD wave functions effectively mitigates signal loss at large transverse separation, enabling reliable nonperturbative extraction up to $b_\perp\sim 1{\rm fm}$. The results for CS kernel extracted via this automated approach  serve as critical input for global TMD fits, and the framework can be readily extended to additional lattice ensembles and other TMD observables with slight modifications.

We note, however, that the generation of raw lattice QCD data is not addressed in the present work, with \textsc{PhysMaster} taking existing lattice correlator and Wilson-loop data as input. While the parametrization given in Eqs. (9) and (10) is physically motivated, it introduces model dependence, and our current study is limited by the lack of explicit transverse-momentum dependence in the adopted parametrization, as well as statistical uncertainties inherited from the input lattice data.

To address these limitations, we will advance this AI-automated lattice QCD paradigm along several key directions in future work. We plan to incorporate additional lattice configurations and finer lattice spacings to further reduce statistical and systematic uncertainties, extend the automated framework to nucleon TMD observables for realistic phenomenological applications, and integrate neural network-based parametrizations to capture complex nonperturbative behavior without manual model design. Most importantly, we will expand \textsc{PhysMaster} to incorporate the full generation of lattice QCD data in future developments, moving toward a fully end-to-end autonomous workflow.

As an AI co-scientist, \textsc{PhysMaster} automates labor-intensive numerical routines while preserving full physical interpretability, freeing researchers to focus on physical interpretation and theoretical innovation. This work establishes a reproducible, scalable blueprint for autonomous AI-driven studies of parton structure and other non-perturbative observables in lattice QCD.

\section*{Acknowledgments}
We thank LPC for the collaboration of Ref.~\cite{Tan:2025ofx}.   This work is supported in part by Natural Science Foundation of China under Grants  No. 12125503, No. 12305103,

\newpage
\appendix
\section{The Architecture of \textsc{PhysMaster}}
\label{physmaster}
\textsc{PhysMaster} is a multi-agent system designed to operate as an autonomous theoretical and computational physicist~\cite{Miao:2025sms}. The system integrates theoretical reasoning with executable numerical computation and is structured to support ultra-long-horizon scientific workflows.

At the beginning of the workflow, the clarifier transforms the original natural-language query into a structured research task.  More expliictly, the clarifier extracts essential information from the query and formulates it into a structured contract, including the research topic, domain, task description, expected input and output formats, and relevant physical constraints such as symmetries, conservation laws, dimensional analysis, and characteristic scales. In addition, the clarifier decomposes the problem into a sequence of executable subtasks with and tractable intermediate results that can be dynamically scheduled during later stages of the workflow. 

\begin{figure}[htb]
\centering
\includegraphics[width=0.8\linewidth]{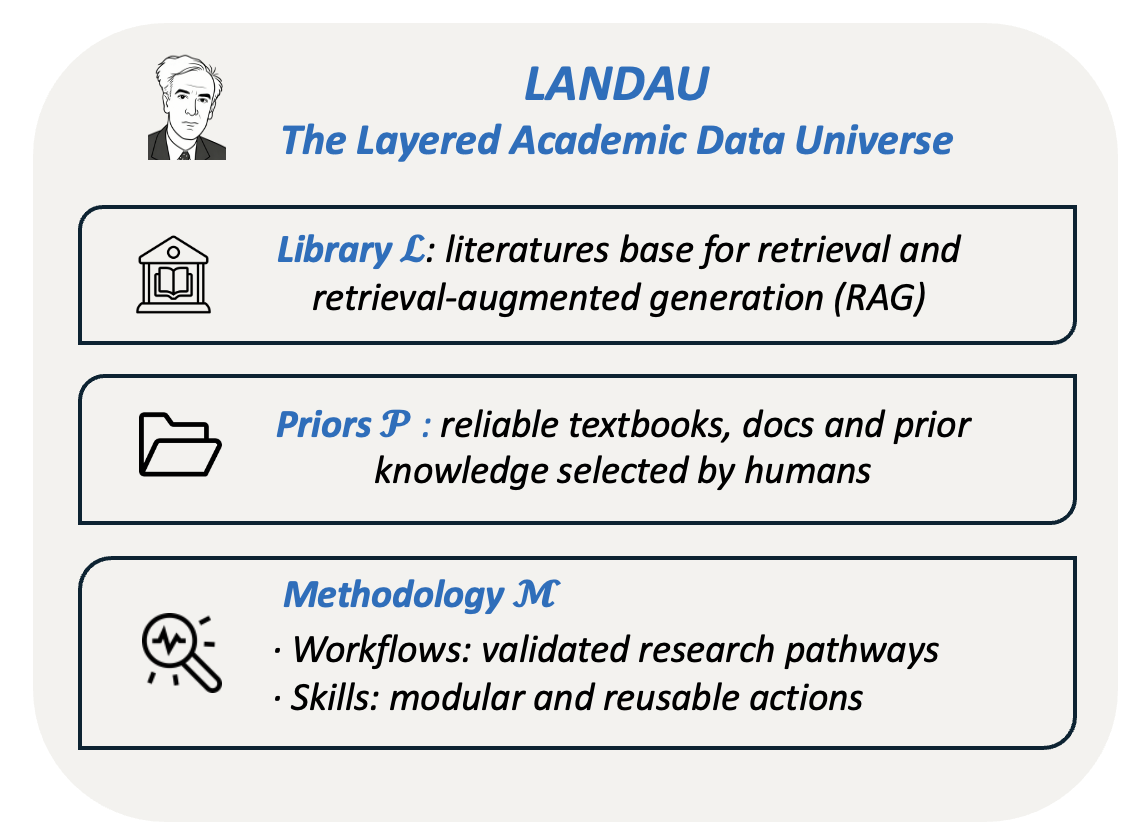}
\caption{ Structure of the LANDAU layered academic data universe, consisting of literature library, validated methodologies, and high-confidence physics priors to support reliable AI-automated theoretical and computational physics research.}
\label{fig:landau}
\end{figure}

To provide reliable background knowledge for the autonomous research  process, \textsc{PhysMaster} is equipped with a persistent knowledge infrastructure \textsc{LANDAU}, the Layered Academic DAta Universe, which is named in honor of the eminent physicist Lev Landau.  
As shown in Fig.~\ref{fig:landau}, the \textsc{LANDAU} system organizes scientific knowledge into three major components:
\begin{eqnarray}
{\rm LANDAU } = \mathcal{L} \cup \mathcal{M} \cup \mathcal{P},
\end{eqnarray}
where $\mathcal{L}$ denotes the library of knowledge extracted from retrieved scientific papers, $\mathcal{M}$ represents the validated methodology consisting of effective reasoning paths and technical workflows, and $\mathcal{P}$ contains high-confidence priors manually selected from textbooks or authoritative sources.  Together, these layers provide both conceptual guidance and quantitative references for subsequent reasoning and computation. 

The core problem-solving process is carried out during the task-execution stage through a hierarchical agent collaboration framework combined with Monte Carlo Tree Search (MCTS) as inspired by Ref.~\cite{Liu:2025brr}.

\begin{figure}[htb]
\centering
\includegraphics[width=0.8\linewidth]{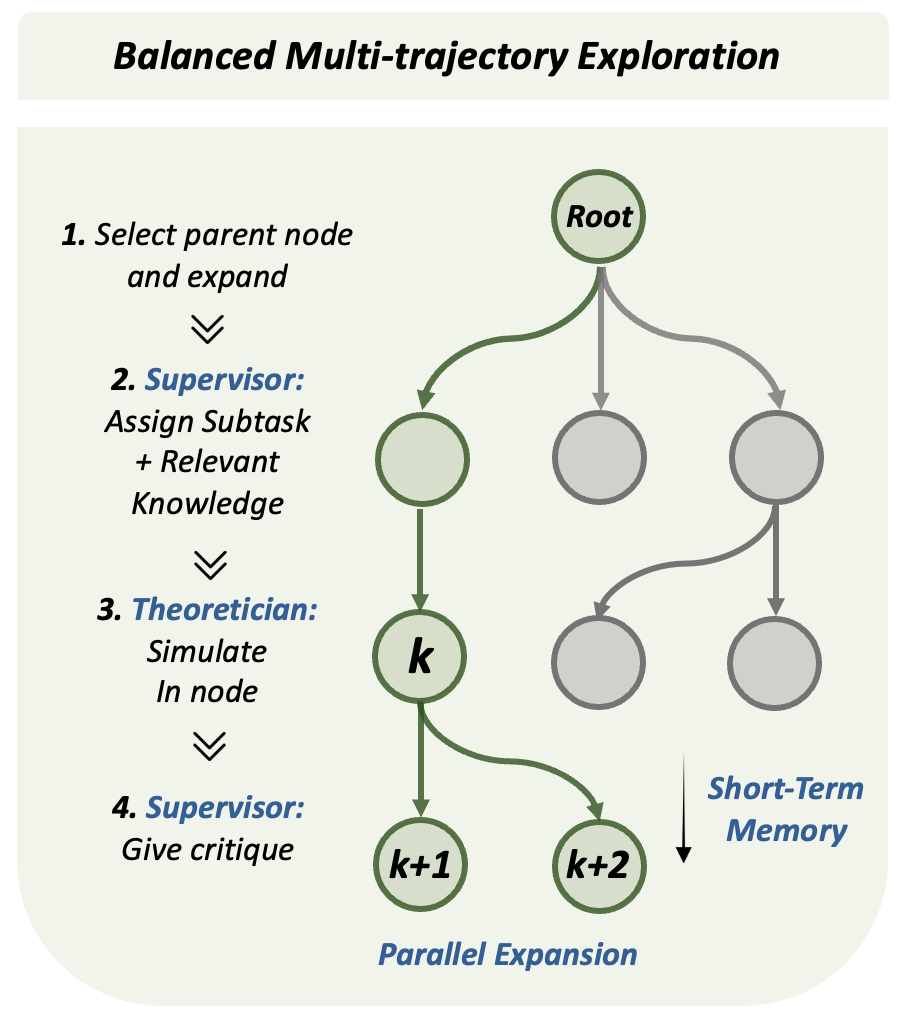}
\caption{The Monte Carlo Tree Search (MCTS) exploration of \textsc{PhysMaster}.}
\label{fig:MCTS}
\end{figure}

Real research problems in theoretical and computational physics typically require extensive iteration of modeling, derivation, coding, and verification. Such problems often require externalized progress management and exploration of multiple candidate solution trajectories, as shown in Fig.~\ref{fig:MCTS}. In \textsc{PhysMaster}, each node in the MCTS search tree represents an attempt to construct or refine a partial solution.  Subtasks generated by the clarifier are assigned to these nodes, together with concise summaries of previous exploration results and relevant background knowledge retrieved from \textsc{LANDAU}. 
The search process expands multiple trajectories in parallel, enabling exploration of diverse modeling strategies or computational implementations.

During this stage, two agents interactively   collaborate:
\begin{itemize}
    \item The ``Supervisor" serves as both a scheduler and a critic. It manages global scheduling, assigns subtasks to be executed at each node and provides the necessary background knowledge from \textsc{LANDAU}. Meanwhile, it evaluates intermediate results using evidence retrieved from \textsc{LANDAU} and assigns scalar rewards to nodes, summarizes the current exploration state, and generates actionable critiques that guide subsequent refinement steps. 
    \item The ``Theoretician" is responsible for executing the assigned subtask. Depending on the nature of the task, it may construct theoretical models, perform analytical reasoning, derive equations, or translate the model into executable code for numerical computation. 
\end{itemize}

Reliable feedback is essential for guiding long-horizon exploration. 
The signals given by supervisor are fed into the tree policy to determine future node selection and expansion. 
Nodes are selected according to the UCT (Upper Confidence bounds applied to Trees) criterion
\[
\mathrm{UCT}(v) = \frac{Q_v}{N_v} + C\sqrt{\frac{\ln N_{\mathrm{parent}}}{N_v}},
\]
where $Q_v$ and $N_v$ denote the accumulated reward and visit count of node $v$, respectively, and $C$ controls the trade-off between exploration and exploitation. 
Through this structured exploration process, the system gradually identifies high-quality solution trajectories while continuing to explore alternative approaches.

To conclude, \textsc{PhysMaster} integrates structured task formulation, a layered scientific knowledge infrastructure, and MCTS-based hierarchical reasoning to support ultra-long-horizon research workflows, combining theoretical reasoning with executable computation. This architecture enables autonomous, PhD-level independent exploration across research domains reliant on rigorous formal theoretical analysis and advanced numerical computations, including high-energy and particle theory, condensed matter theory, cosmology and astrophysics, and quantum information. In turn, \textsc{PhysMaster} establishes a collaborative paradigm between physicists and AI agents: physicists provide the physical insight, while the AI undertakes the full execution of labor-intensive computational and analytical tasks, yielding a substantial enhancement in overall research efficiency.

\clearpage
\bibliographystyle{apsrev4-1} 
\bibliography{main_ref} 

\end{document}